\documentclass{iaus}
\usepackage{graphicx}
\fulloutfootnote

\title[The VO and the Time Domain] 
{Using the VO to Study the Time Domain}

\author[ R. Seaman, R. Williams, M. Graham \& T. Murphy ]   
{Rob Seaman$^1$, Roy Williams$^2$, Matthew Graham$^2$, \and Tara Murphy$^3$}

\affiliation{$^1$National Optical Astronomical Observatory, Tucson AZ USA \\
email: {\tt seaman@noao.edu} \\[\affilskip]
$^2$California Institute of Technology, Pasadena CA USA \\
email: {\tt mjg@caltech.edu, roy@caltech.edu} \\[\affilskip]
$^3$University of Sydney, Australia \\ 
email: {\tt tara@physics.usyd.edu.au}}

\pubyear{2011}
\volume{285}  
\pagerange{1--6}
\jname{New Horizons in TIme Domain Astronomy}
\editors{R.E.M. Griffin, R.J. Hanisch, \& R. Seaman, eds.}
\begin{document}

\maketitle

\begin{abstract}
Just as the astronomical ``Time Domain" is a catch-phrase for a diverse group
of different science objectives involving time-varying phenomena in all
astrophysical r\'egimes from the solar system to cosmological scales, so the
``Virtual Observatory" is a complex set of community-wide activities from
archives to astroinformatics.  This workshop touched on some aspects of
adapting and developing those semantic and network technologies in order to
address transient and time-domain research challenges.  It discussed the
VOEvent format for representing alerts and reports on celestial transient
events, the SkyAlert and ATELstream facilities for distributing these alerts,
and the IVOA time-series protocol and time-series tools provided by the VAO.
Those tools and infrastructure are available today to address the real-world
needs of astronomers.  

\end{abstract}

\firstsection 
\section{Introduction}
Virtual Observatory activities relating to time-domain astronomy have focused on
the VOEvent celestial transient alert protocol since 2005.\footnote{IVOA
VOEvent Working Group Wiki: {\tt http://voevent.org}} VOEvent is a message format for
conveying reports of time-varying astronomical observations (\cite{seaman11}).
The notion of VOEvent is to permit the several different pre-existing
astronomical telegram standards and transient alert protocols to be expressed
in a common form that will permit interoperability, while enabling modern
virtual technologies to be used to construct autonomous workflows.  The goal is
to close the observational loop from the discovery of transient phenomena, for
instance by large synoptic surveys, to their follow-up by robotic or
human-mediated instrumentation and telescopes.

VOEvent is necessary but not sufficient.  A common message format requires an
interoperable transport infrastructure layer.  Three of these---SkyAlert,
ATELstream and the VAO Transient Facility (\pageref{VO_Workshop})---are
discussed below.  Others such as the VOEvent-enabled Gamma-ray Bursts
Coordinates Network (GCN) and the IAU CBAT service (Central Bureau of
Astronomical Telegrams) have also been adapted to VOEvent compliance.
VOEvent, like any living standard, must also evolve to meet the needs of the
community, so the recent VOEvent v2.0 Recommendation (standard) of the
International Virtual Observatory Alliance (IVOA, an activity of Commission 5
of the IAU) is also discussed below.

A key aspect of time-domain astronomy is the collection and interpretation of
time-series data sets.  This is related to VOEvent, but is also a key area for
IVOA-compliant data archives.  Recent work on time-series tools, performed by
the U.S. Virtual Astronomical Observatory (VAO), was demonstrated during the
workshop.

\section{Skyalert}
Skyalert\footnote{http://skyalert.org} is a clearing-house and repository of
information about astronomical transients, each described by a collection of
VOEvent packets that may have multiple authors. The components of Skyalert
are:
\vskip6pt
\begin{itemize}
\item
A Web-based event broker, allowing subscription so that information about
transients can be delivered to users and their telescopes immediately upon
receipt.

\item 
A Web-based authoring system, so that authenticated users can inject events
direct from automated discovery pipelines, or fill in Web forms, that may be
delivered rapidly to others.

\item 
An event repository, storing all events that come through the broker, and
allowing bulk queries and drill-down.

\item 
A \emph{click or code } paradigm that allows people Web-based access and
machines Web-service access.

\item 
A way to see recent and past transients: as tables, multi-layered Web pages, or
with popular astronomical software.

\item 
A development platform for building real-time decision rules about transients,
and for mining the repository.

\item 
Open-source software to allow local implementations as well as the Web-based
application.
\end{itemize}

\vskip6pt
The crucial standard that enables interoperable exchange of events is called
VOEvent, now a Recommendation of the International Virtual Observatory
Alliance. Reading that standard, as described above, is a good basis for
understanding more of this document.  Skyalert installation shows on the front
page all the recent events (last 200) ingested into the system, as clickable
dots in a semi-log timescale, with the present moment at the right, and older
events further top the left. Clicking on any of the dots brings up the
portfolio for that event. Also available from the front page is a collection of
Atom feeds of recent events (both system feeds and your custom feeds).

\subsection {Event Portfolios}

Each transient will have a collection of data that we call a data portfolio:
a collection of numbers, links, images, opinions, search results, etc.  A
portfolio is defined through a citation mechanism inherent in the VOEvent
packet, where one event can cite another. Thus, an event with no citation
becomes its own portfolio, but an event with a citation to another joins the
portfolio to which the other belongs.

As noted above, the portfolio detail page can be accessed by clicking on the
dot for a recent event. One can also select a specific event stream via
\emph{Browse Event Streams}, choose the required table of portfolios, and then
narrow the search to a specific event. A third route to get to a specific event
is by selecting a feed (one's own or a system feed), and then choosing a
portfolio at that point.  There are three representations that are available
for each event of a portfolio:

\begin{itemize}

\vskip6pt
\item Overview: created by running the event data through the overview template.

\item Params: a table of parameters and their values, plus representation of any
Tables in the event. 

\item XML: showing the actual XML that was loaded into Skyalert

\end{itemize}

\subsection{Alerts}
An alert is a means of determining whether a portfolio is ``interesting'' in
some way, and what to do if it is. From each ``rule'' (see below) is
automatically generated a feed of interesting portfolios. It may be that the
action which results from an interesting portfolio can cause another event to
be loaded to the same portfolio, and that might in turn cause another rule to
be satisfied, and another action to be taken. 

Each alert has a collection of streams; for the rule to operate on a portfolio,
it must have an event drawn from each of its needed streams. The simplest
rules, however, need only events from one stream; for example, an
event from a stream called {\it apple} might be interesting if it is bright; if 
there is a Param named {\it magnitude} then the trigger expression might be:
\vskip6pt
\begin{verbatim}
     apple[magnitude] < 18
\end{verbatim}
\vskip6pt
The trigger expression is interpreted by Python, within
a sandbox environment that allows only math.~functions.  Thus, a trigger
expression like {\tt os.system(rm *)} will fail because the {\tt os} module
cannot be imported into the sandbox environment.

A rule runs on a collection of events (i.e.~a portfolio). The simplest rule
considers only one stream.  Rules can be created only by a user who is
registered and logged in.  To build a rule, the user first clicks on \emph{my
feed and alerts}, then on \emph{for a new alert}, and then selects an event
stream (but should not click the advanced option).  A name for the alert is
entered, to start to make the trigger expression. The simplest expression is
\emph{True}, meaning that all of the events are interesting. The user must
explicitly save the rule before the latter can do anything; the expression is
checked for syntax before being saved, so that only syntactically correct
trigger expressions get into the database.  The alert-editor screen also has a
button to show all the past events that would have satisfied the trigger.  More
complex rules can use multiple event streams to make a joint decision on the
portfolio; suppose (for example) that an event from the {\it apple} survey has
a follow-up from the {\tt fruitObserver} catalogue, and that bright apple
events are required which are also bright in the {\tt fruitObserver} stream:

\vskip6pt
\begin{verbatim}
    apple[magnitude] < 18 and fruitObserver[gMag] < 18
\end{verbatim}
\vskip6pt

This a joint criterion on two different event streams, authored
by different people. In one stream the author chose to use the Param called
\emph{magnitude}, and in the other stream the author chose to use the name
\emph{gMag}. Because of the underlying VOEvent model, Skyalert is able to
integrate information from multiple authors.

\subsection{Layering facilities on VO-compliant protocols: VOEvent2}
The IVOA VOEvent standard was first defined, and a prototype created, in 2005,
and reached official Recommendation status in 2006.  It has been adopted
successfully by many astronomical time-domain projects since then.  It was
recognized from the start that more advanced features would be needed in order
to grapple with the challenging time-domain projects looming in the near
future.  The recently-adopted VOEvent2 standard embraces such features.

\subsection{VO compliant protocols: Distributed Transient Facility}

The time-domain community wants robust and reliable tools to enable the
production of, and subscription to, community-endorsed event notification
packets (VOEvents). The proposed Distributed Transient Facility (DTF) is being
designed to be the premier brokering service for the community, not only
collecting and disseminating observations about time-critical astronomical
transients but also supporting annotations and the application of intelligent
machine learning to those observations. Two types of activity associated with
the facility can therefore be distinguished: core infrastructure, and user
services.  The prior art in both areas were reviewed by the workshop, and
planned capabilities of the DTF were described. In particular, it focused on
scalability and quality-of-service issues required by the next generation of
sky surveys, such as LSST and SKA.

\section{ATELstream}

Bob Rutledge of the Astronomer's Telegram described to the workshop the new
ATELstream\footnote{\tt http://blogs.astronomerstelegram.org/atelstream/}
facility that provides a UNIX socket-based XML-driven messaging service for
celestial transient event notices.  Much interest was exhibited in ensuring
that ATELstream and VOEvent remain interoperable and that the community meld
together the two technologies for the benefit of all.

\section{VOEvent2}

This workshop was only the latest in a long line of workshops, meetings and
sessions over the past half-dozen years concerned with using Virtual
Observatory standards and protocols for studying the time domain.  All have
been tied to VOEvent, but also come more generally under the banner of
\emph{Hot-wiring the Transient
Universe}\footnote{\tt http://hotwireduniverse.org/}.  Copies of the same-named
book (\cite{hotwired}) were made available for distribution.

The system architecture of the IVOA consists of numerous protocols, data models
and services.  The VOEvent Recommendation is one of the many diverse
international standards of the IVOA.  VOEvent refers to, and relies on, several
of those, and other standards may in turn depend on VOEvent.  VOEvent is also
engaged in numerous external projects; several are currently distributing
messages in VOEvent format, others are connected to major future surveys, and
still others represent existing projects with a stake in enhancing their
interoperability.

Planning is underway for a third-generation transport infrastructure.  The
diverse prototype technologies of the original VOEventNet were followed by the
deployment of the operational SkyAlert system discussed above.  Future
celestial transient event transport infrastructure within the Virtual
Observatory is anticipated to develop from the VAO transient facility project
described below.  In the mean time, numerous efforts will continue at working
with other transient alert technologies and projects to ensure interoperability
both across the community and for all types of time-varying celestial
phenomena.  Building such an infrastructure is an exercise in creative
bootstrapping.

The IVOA VOEvent standard became a Recommendation of the IVOA in 2006.
Evolution of the format was planned, and a major milestone was reached in 2011
with the acceptance of the VOEvent v2.0 update to the standard.  Its major
features include:

\vskip6pt
\begin{itemize}
\item 
The definition of VOEvent \emph{Streams} in support of ongoing work to enhance
the registration of VOEvent resources in the Virtual Observatory.

\item 
VOEvent has always been transport-neutral.  In recognition of the
rapidly changing landscape of transport technologies, the explicit description
of transport options has been removed from the standard.

\item 
The VOEvent \emph{$\langle$Param$\rangle$} element has been generalized.

\item 
A facility was added to embed general-purpose tabular data structures within
VOEvent packets.

\item 
Time series will be supported as tables with \emph{utypes} referencing an IVOA
time-series data model.

\item The VOEvent \emph{$\langle$Reference$\rangle$} element has been generalized.
\end{itemize}
\vskip6pt

In addition to changes to the VOEvent format, the v2.0 efforts focused on
creating a more robust XML schema and on enhanced libraries and Web access
compatible with the format and schema.  The fundamental nature of the VOEvent
format has remained the same between versions.  VOEvent exists to support the
engineering of empirical workflows (\cite{seaman08}) and with
key elements:
\vskip6pt
\begin{itemize}
\item \emph{$\langle$Who$\rangle$} -- author's provenance
\item \emph{$\langle$What$\rangle$} -- empirical measurements
\item \emph{$\langle$WhereWhen$\rangle$} -- targeting in spacetime
\item \emph{$\langle$How$\rangle$} -- instrumental signature
\item \emph{$\langle$Why$\rangle$} -- scientific characterization
\item \emph{$\langle$Citations$\rangle$} -- building threads of transient-response follow-up
\end{itemize}

\section{VAO Time-Series Tools}

Systems to disseminate event notifications are a major component of the VO's
infrastructure for supporting time-domain science. Equally as important,
however, are the tools and services that enable and facilitate the discovery
and analysis of collections of time-series data.  In view of the increasing
number of new time-domain surveys now in progress or being planned, providing a
framework to interconnect the data in distributed archives and appropriate
services can only aid both the discovery of previously unknown phenomena and
improve our understanding of already known ones.  Through a number of
activities, the VAO aims to create such interoperability, allowing astronomers
to locate the data they want and then effortlessly connect the data providers
to different types of available tools, such as a periodogram service or a
time-series modeller, as part of a workflow or scripted analysis session.

\begin{figure}[h]
\begin{center}
\includegraphics[width=5.0in]{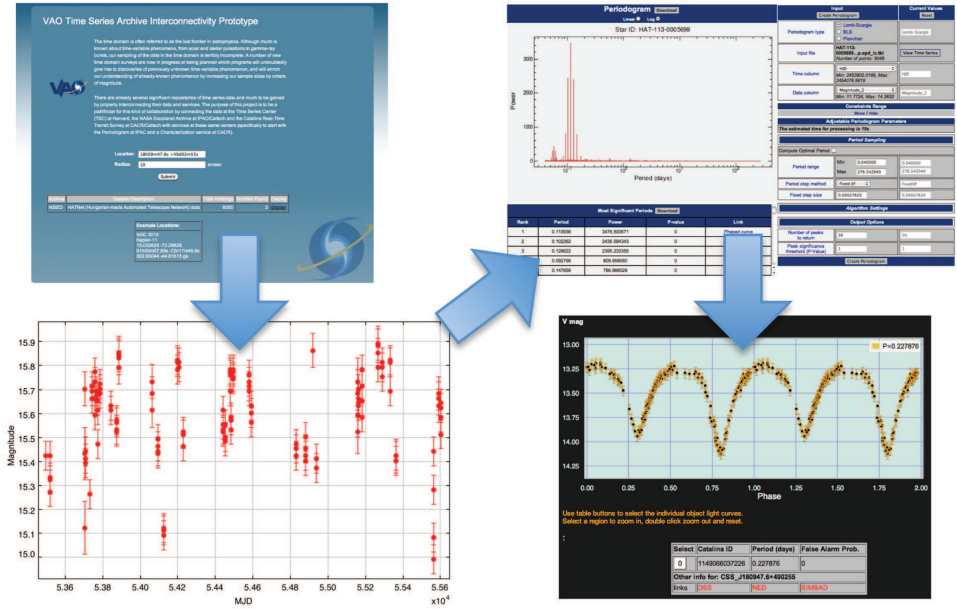} 
\caption{Time-series data workflow facilitated by the VAO.}
\label{fig1}
\end{center}
\end{figure}

As an illustration of such a system, one might consider an astronomer who is
using variable stars, say RR Lyrae types, to study Galactic structures such as
spiral arms, stellar streams or the like. A major contaminant in that kind of
analysis can be eclipsing binaries, and the usual light curves of both classes
can be difficult to distinguish.  However, the binaries can easily be filtered
out using phased light curves. The astronomer could therefore create a pure
data set for the analysis by identifying suitable data through the VAO Time
Series Archive Interconnectivity Portal and sending them to the periodogram
service at the NASA Exoplanet Archive. The VAO infrastructure will handle the
data transfer for the astronomer and the result that is returned is a phased
light curve (see Fig.~\ref{fig1}).

A pathfinder for this kind of collaboration is being developed by the VAO,
initially connecting the Harvard Time Series Center, the NASA Exoplanet Archive
and the Catalina Real-Time Transient Survey. As more data sets and tools become
available, they will be integrated seamlessly. Provision for bulk activities,
such as the large-scale characterization of time series, is also being
considered.

\section{Looking Ahead}

During the workshop a wide-ranging discussion followed (and often interrupted)
the presentations.  There appeared to be a strong consensus that not only would
the time domain increase in importance for astronomy in the future, but that to
take advantage of its full potential virtual and semantic technologies would be
critical.  The often lively discussion addressed future directions for the
virtual observatory time-domain facilities as a whole.

\end{document}